\documentclass[twocolumn,preprintnumbers,amsmath,amssymb,prl]{revtex4}

\usepackage{graphicx}
\usepackage{dcolumn}
\usepackage{bm}
\begin{document}


\title{Point contact spectroscopy of Pr$_{2-x}$Ce$_x$CuO$_4$ in high magnetic fields}

\author{Sung Hee Yun$^1$}
\author{Neesha Anderson$^1$}%
\author{Bing Liang$^2$}
\author{R. L. Greene$^2$}
\author{Amlan Biswas$^1$}
\affiliation{$^1$Department of Physics, University of Florida,
Gainesville, Florida 32611, USA\\
$^2$Center for Nanophysics and Advanced Materials, Department of
Physics, University of Maryland, College Park, Maryland 20742, USA}

\date{\today}

\begin{abstract}
We have studied the density of states of the normal state of the
electron-doped superconductor Pr$_{2-x}$Ce$_x$CuO$_4$ at low
temperatures and in magnetic fields up to 31 Tesla using point
contact spectroscopy on single crystals. Our data clearly reveal an
anomalous gap in the normal state density of states. This normal
state gap survives even in the highest applied field of 31 T. Our
results cast doubt over whether this gap found in electron-doped
cuprates is the analog of the pseudogap in hole-doped cuprates. We
have suggested an alternate origin of the normal state gap, which
involves the effect of disorder and electron correlations at the
surface of cuprates.

\end{abstract}

\pacs{Valid pacs appear here}
 \maketitle
The normal state of copper-oxide (cuprate) superconductors has been
the subject of intense research because it may hold the key to
understanding the mechanism leading to the observed high-critical
temperatures ($T_c$) in these materials. In conventional
superconductors both the formation and condensation of paired
carriers into a zero resistance state occurs at the critical
temperature $T_c$, which is accompanied by the formation of the
superconducting gap. A striking feature of hole-doped cuprates is
the formation of a gap in the density of states well above $T_c$,
{\em viz.} the pseudogap ~\cite{renner,yazdani,timusk}. Several
models have been suggested to explain the origin of the pseudogap,
which can be sorted into two main categories {\em viz.} the
pre-formed pairing theory ~\cite{emery,randeria} and the competing
order parameter theory ~\cite{varma,chakravarty}. Preformed pairing
theory considers the pseudogap state to be a phase fluctuation
regime which is a precursor to the superconducting state whereas,
the competing order parameter models suggest that the pseudogap is
due to another competing phase.

The pseudogap has been studied extensively in hole-doped ($p$-doped)
cuprates mainly by probing the region outside the superconducting
dome in the temperature - hole-doping ($T - x$) phase
diagram~\cite{timusk}. The presence of a pseudogap in electron-doped
($n$-doped) cuprates is more controversial. One example is the
difference in the energy scales of the gaps measured in the normal
state of $n$-doped cuprates using different techniques. While
tunneling and point-contact spectroscopy show a normal state gap in
$n$-doped cuprates with an energy scale similar to the
superconducting gap ($\sim$ 10meV)~\cite{alff1,biswas,qazilbash},
angle-resolved photoemission spectroscopy and optical reflectivity
reveal a much larger gap ($\sim$ 100 - 200 meV), which is attributed
to a spin-density wave gap ~\cite{armitage,onose}. In the normal
state of $p$-doped cuprates there is reasonable agreement among the
gaps measured using different experimental techniques~\cite{timusk}.
[Note: We refer here to the higher energy gap found in the normal
state of $p$-doped cuprates which increases with decreasing doping
and not the lower energy gap which decreases with decreasing doping
as recently reported in ref.~\cite{tanaka}. For the rest of this
manuscript we will refer to the higher energy gap in $p$-doped
cuprates, as the pseudogap.] Furthermore, the phase fluctuation
region above $T_c$ observed by Nernst effect measurements in
$p$-doped cuprates is either absent ~\cite{ong} or is smaller in
$n$-doped cuprates ~\cite{pengcheng}.

A unique feature of the $n$-doped cuprates is that it is possible to
probe the normal state {\em inside} the superconducting dome for the
entire doping range by driving the superconductor into the normal
state with a field $H$ greater than the upper critical field
$H_{c2}$, due to the relatively low values of $H_{c2}$ ($\sim$ 10 T
at low temperatures for the optimally doped compounds). Tunneling
across grain boundary junctions and point-contact spectroscopy
results in magnetic fields $H > H_{c2}$ have clearly shown the
presence of a normal state gap (NSG) in the density of states, which
is comparable in energy scale to the superconducting gap and which
increases in energy scale as the doping is
decreased~\cite{alff1,biswas,qazilbash}. Alff {\em et al.} have
mapped the region in the phase diagram where this NSG is
observed~\cite{alff2}. The NSG observed in $n$-doped cuprates shows
a similar behavior with doping as the high energy pseudogap in
$p$-doped cuprates~\cite{alff2,biswas,tanaka}. If the NSG in
$n$-doped cuprates is indeed analogous to the pseudogap in $p$-doped
cuprates, then Alff {\em et al.} result suggests that the pseudogap
in $n$-doped cuprates is present inside the superconducting dome and
vanishes at a doping level close to the optimum value ($x \sim
0.16$), which supports the competing order parameter
scenario~\cite{alff2}. However, Dagan {\em et al.} have performed
tunneling spectroscopy measurements on thin films of
Pr$_{2-x}$Ce$_x$CuO$_4$ for the doping range $0.11 < x < 0.19$ and
have shown that the NSG is observed well into the overdoped
region~\cite{dagan}. In fact, in the overdoped region ($x
> 0.17$) the temperature $T^*$ above which the NSG vanishes is
approximately equal to $T_c$. This result provides strong evidence
that the NSG is related to the superconductivity in PCCO and if the
NSG is the pseudogap, supports the pre-formed pairing scenario.
Recently, Kawakami {\em et al.} have reported the observation of a
pseudogap in the $n$-doped cuprate Sm$_{2-x}$Ce$_x$CuO$_{4-\delta}$
using interlayer tunneling transport~\cite{kawakami}. By studying
this pseudogap's behavior in magnetic fields up to 45 T, Kawakami
{\em et al.} concluded that the pseudogap in both $n$-doped and
$p$-doped cuprates is formed due to spin-singlet (pair) formation
above $T_c$~\cite{kawakami}. By comparing the behavior of the NSG in
high magnetic fields to the pseudogap reported in
ref.~\cite{kawakami}, we expect to obtain clues to explain the
origin of the NSG. In this paper we report the first point contact
spectroscopy (PCS) measurements on single crystals of optimally
doped Pr$_{2-x}$Ce$_x$CuO$_4$ ($x = 0.15$, PCCO) in magnetic fields
up to 31 T. We show that the NSG survives even in a field of 31 T,
which casts doubt over the validity of the assumption that the NSG
is the pseudogap in $n$-doped cuprates. We suggest alternate origins
for the formation of the NSG.

Point contact spectroscopy (PCS) is similar to scanning tunneling
spectroscopy, in the sense that the current injection occurs between
a sharp tip and the sample. However, in PCS the tip is actually in
physical contact with the sample. Hence, PCS is less susceptible to
mechanical vibrations than scanning tunneling spectroscopy, which is
an important factor to consider when choosing a measurement method
for the water-cooled magnets in the DC High Field Facility at the
National High Magnetic Field Laboratory in Tallahassee (NHMFL). The
PCS data were taken using a custom built probe designed for
operation in the 32 mm bore DC field magnets at the
NHMFL~\cite{nhmfl}. The maximum magnetic field is 33 tesla at
temperatures down to 1.5 K. To tunnel into the $a-b$ plane and also
apply a magnetic field perpendicular to the $a-b$ plane (see Fig. 1,
right inset) we used bevel gears to suitably change the direction of
tip-sample approach (Fig. 1 inset). For finer control over the
junction resistance, we used a worm-gear arrangement (1:96 ratio) at
the top of the probe. Differential conductance (d$I$/d$V$ $\equiv
G$) vs. $V$ curves were obtained directly by using a modulation
technique. Details of the apparatus and electronics are given in
ref.~\cite{nhmfl}. The PCCO crystals were grown by the self-flux
technique followed by an oxygen reduction procedure to achieve a
$T_c \approx 21 \pm 0.8$ K~\cite{balci}. The data presented in this
paper were taken for different point contact junctions on the same
single crystal. In addition, we have confirmed the presence of the
pertinent density of states features at high magnetic fields for
another single crystal and a thin film of PCCO.

\begin{figure}[t]
\includegraphics[width=8cm,height=6cm]{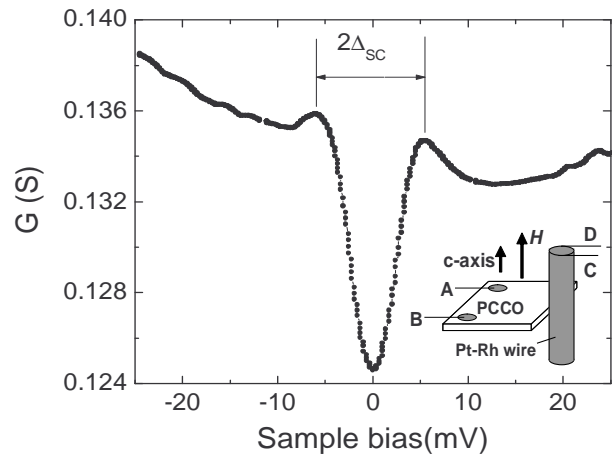}\\
\caption{Differential conductance (d$I$/d$V$ $\equiv G$) - $V$ curve
of a point contact junction between Pr$_{2-x}$Ce$_x$CuO$_4$ ($x =
0.15$) and a Pt-Rh wire at $T$ = 1.5 K. The superconducting gap is
marked. The right inset shows the configuration of the point contact
junction, which ensures current injection into the $a-b$ plane. A
small contact area is achieved due to the small radius of the Pt-Rh
wire (0.25 mm) and the 20 $\mu$m thick crystal.}
\end{figure}

Fig. 1 shows a typical d$I$/d$V-V$ curve obtained on a single
crystal of PCCO for $H = 0$. All point contact spectra shown in this
paper were taken at $T$ = 1.5 K. The coherence peaks which are a
signature of the superconducting gap are clearly observed and are
marked in the figure. The large value for d$I$/d$V$ at zero-bias is
due to the high transparency of a point contact junction. In
addition to the coherence peaks, the point contact spectrum also
shows a linear increase in d$I$/d$V$ ($G$) with $V$ for sample bias
above the superconducting gap value. This asymmetric linear
background conductance makes it difficult to quantify the effect of
the magnetic field on the NSG and needs to be removed. We will
describe the procedure for removing the background later in this
report.

\begin{figure}[b]
\includegraphics[width=8cm,height=6cm]{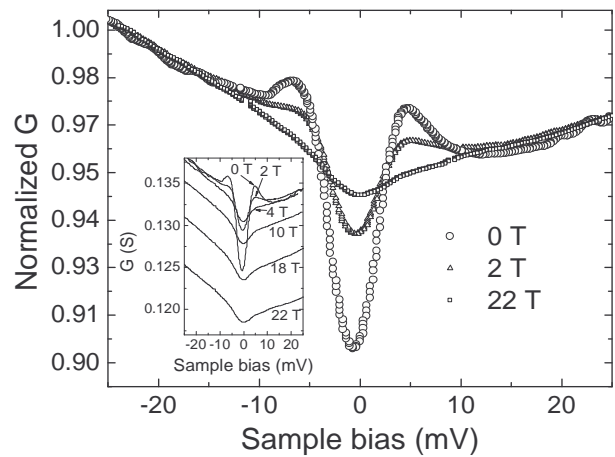}\\
\caption{The effect of a magnetic field on the point contact
spectrum for Pr$_{2-x}$Ce$_x$CuO$_4$ ($x = 0.15$). The normal state
gap is clearly visible in a magnetic field of 22 T. The inset shows
the raw d$I$/d$V$-$V$ curves as a function of the applied magnetic
field.}
\end{figure}

Fig. 2 shows the variation of the $G-V$ characteristics with an
applied magnetic field. Since the resistance of point-contact
junctions is low, the $G-V$ curves shift vertically when the
superconductor becomes normal above $H_{c2}$. Such an effect is also
observed in grain boundary junctions~\cite{alff2}. We have removed
this effect by calculating the change in the resistance of the
sample from the vertical displacement of the $G-V$ curves and also
adjusted the sample bias accordingly~\cite{gvshift}. The normal
state gap is clearly observed even in fields of up to 22 T.

\begin{figure}[t]
\includegraphics[width=8cm,height=12cm]{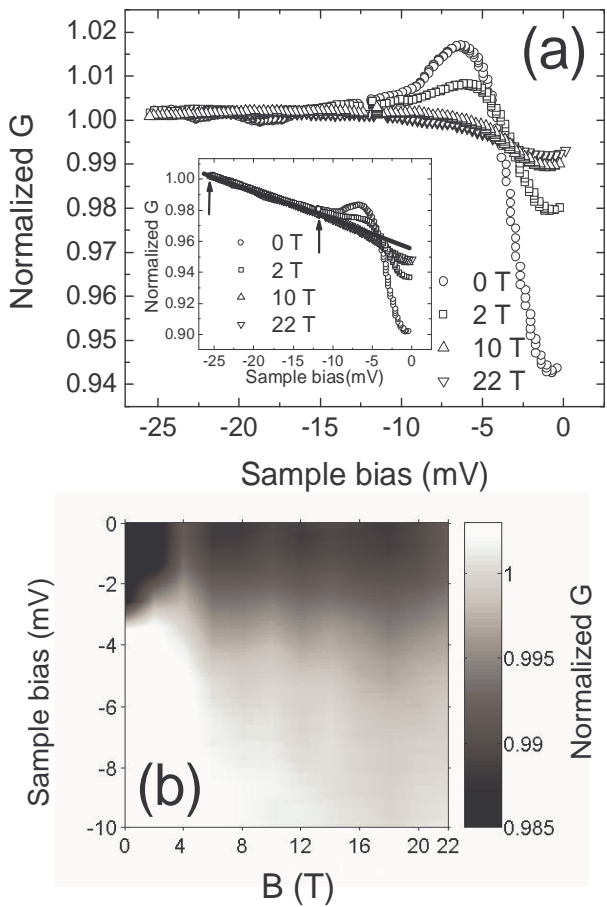}\\
\caption{(a) Normalized point contact spectra showing the effect of
a magnetic field on the normal state gap after a linear background
conductance has been removed from the d$I$/d$V$-$V$ data. The inset
shows the fit to the linear background conductance (solid line),
which is used to normalize the spectra. The arrows indicate the
range for the linear fit. (b) A 2D plot of the point contact spectra
as a function of junction bias and magnetic field, showing the
negligible effect of a magnetic field on the normal state gap.}
\end{figure}

\begin{figure}[t]
\includegraphics[width=8cm,height=12cm]{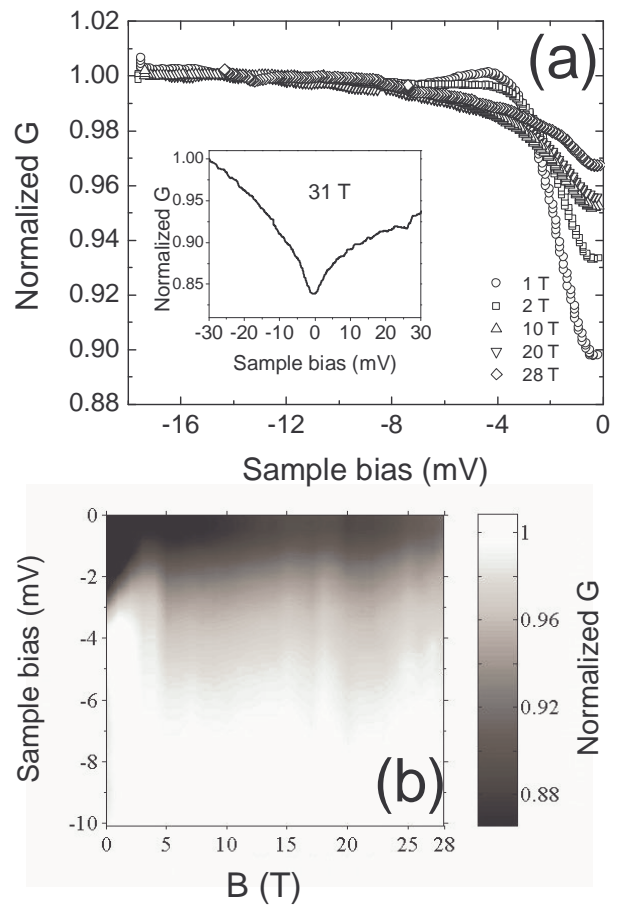}\\
\caption{(a) Normalized point contact spectra in fields of up to 28
T. The inset is a point contact spectrum at 31 T. (b) A 2D plot of
the point contact spectra as a function of junction bias and
magnetic field.}
\end{figure}

\begin{figure}[t]
\includegraphics[width=8cm,height=6cm]{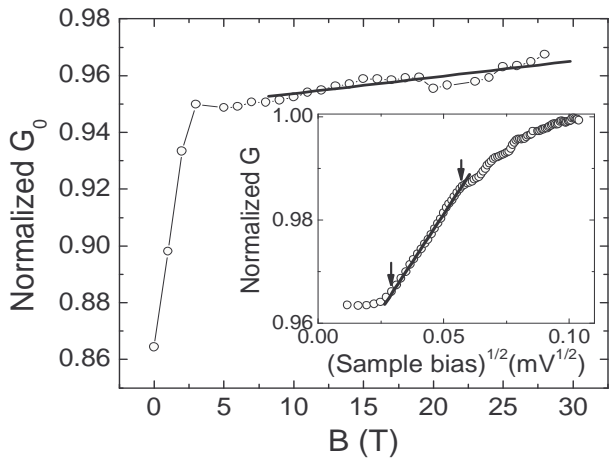}\\
\caption{Variation of the normalized conductance at zero bias as
function of magnetic field. The solid line in a fit to the data from
10 T to 28 T. The inset shows the normalized conductance as a
function of the square root of junction bias. The solid line is a
fit to the data in the range marked by the arrows.}
\end{figure}

To analyze the effect of a magnetic field on the NSG, we have to
remove the linear background and asymmetry of the $G-V$ curves. The
asymmetry of $G-V$ curves has been observed even in scanning
tunneling spectra of cuprates~\cite{pan,yazdani} and has been
explained within the Gutzwiller-Resonating Valence Bond
theory~\cite{andersonong}. However, for our point-contact junctions
and low bias voltages, the origin of the linear background and the
associated asymmetry for the positive and negative biases may also
be due to multiple inelastic scattering through the tunneling
barrier~\cite{kirtley}. Fig. 3 shows our method for background
subtraction which is based on the method suggested by Shan {\em et
al.}~\cite{shan}. Since the background is asymmetric we have used
only the data for negative bias voltages rather than use an
arbitrary background function near zero-bias to connect the
asymmetric linear backgrounds for the positive and negative biases.
The linear background was fitted over the range shown by the arrows
in the inset of Fig. 3. The normalized G-V curves obtained after the
background subtraction is shown in Fig. 3a. The suppression of the
density of states near the fermi level in the normal state of PCCO
is clearly visible even at fields of 22 T. Since the width of the
NSG is difficult to quantify  we show the magnetic field dependence
of the NSG in the form of a 2D plot (Fig. 3b) and it shows that the
magnetic field has a negligible effect on the NSG. Data on a higher
resistance point contact junction is shown in Fig. 4a and shows that
the NSG is present in fields as high as 28 T. We could not obtain
$G-V$ curves at higher magnetic fields without changing the junction
characteristics. However, the normal state gap is still observed in
a field of 31 T as shown in the inset of Fig. 4a. The 2D plot of the
PCS data in Fig. 4a is shown in Fig. 4b and it reveals a small
reduction of the NSG in magnetic fields above 20 T.

From the detailed behavior of the NSG in a magnetic field described
above we now have to answer the question: Is this NSG in $n$-doped
cuprates analogous to the pseudogap observed in $p$-doped cuprates?
Our data clearly shows that the NSG is not suppressed by fields up
to 31 T. A previous report on the pseudogap in $n$-doped cuprates
has shown that the pseudogap closing field, $H_{pg}$ is related to
the pseudogap closing temperature, $T^{*}$ through the Zeeman
equation~\cite{kawakami}:
\begin{equation}
k_BT^{*} = g\mu_BH_{pg}
\end{equation}
Using the value of $T^{*}$ for optimally doped PCCO to be $\approx$
38 K  and equation 1 the corresponding $H_{pg}$ was calculated to be
$\approx$ 28 T~\cite{kawakami}. Hence, the possible reasons for the
survival of the NSG even in fields of upto 31 T are (1) The NSG is
analogous to the pseudogap in $p$-doped cuprates but the $H_{pg}$ is
higher than the value estimated from equation 1, which would suggest
that the origin of the pseudogap is not pre-formed pairs above $T_c$
or (2) The NSG is not analogous to the pseudogap in $p$-doped
cuprates and its origin is unrelated to superconductivity in
$n$-doped cuprates.

To investigate the reason behind the insensitivity of the NSG to
magnetic fields, we have plotted the zero bias normalized
conductance, $G_0$ as a function of magnetic field in Fig. 5. The
initial sharp rise in $G_0$ from 0 to about 3 T is due to the
suppression of the superconducting gap. For $H
> 3$ T, the gradual rise in $G_0$ indicates the slow reduction of
the NSG with field. Assuming the simplest function for the variation
of $G_0$ with $B$, we have fitted a line to the data. From the
linear function we estimate that $G_0$ will approach 1 (which
signifies the complete suppression of the NSG) at a magnetic field
of $\approx$ 90 T. The energy associated with this value of the
magnetic field is $g\mu_BH \approx$ 10 meV, which is comparable to
the width of the NSG. Although the above estimate is crude, such a
correspondence of the energy scales is reminiscent of the effect of
a magnetic field on the zero bias anomaly (ZBA) formed due to
electron-electron interactions in the density of states of
disordered metals~\cite{imry,altshuler}. In 3-D the ZBA has the form
$n(E) = n(0)(1 + \sqrt{E/\Delta})$, where $\Delta$ is called the
correlation gap. The inset of Fig. 5 shows a plot of $G$ as a
function of the square-root of the bias voltage in a magnetic field
of 26 T. Although the linear behavior (solid line) of $G$ in the
bias range 0.8 mV $< V <$ 3.5 mV suggests that the origin of the NSG
could be electron-electron interactions, there are two problems with
this conclusion. First, the ZBA or the correlation-gap is observed
in materials which show a ln($T$) conductivity whereas Dagan {\em et
al.} have shown that the NSG exists even in overdoped PCCO which
shows a metallic conductivity~\cite{dagan}. However, it is possible
that the NSG is purely a surface phenomenon since it has been
observed only in tunneling and point contact experiments on
PCCO~\cite{alff1,biswas,qazilbash,alff2}, which are surface
sensitive probes~\cite{fisher} and it is suspected that the surface
of cuprates can sometimes show properties different from the
bulk~\cite{loram}. In addition, such features have also been
observed in tunneling spectra of other metallic
oxides~\cite{raychaudhuri}. Secondly, the scaling of the width of
the NSG with the superconducting gap (as a function of doping),
strongly suggests that the NSG is related to superconductivity
although, it is also possible that the evolution of the NSG with
doping is related to the insulator-to-metal crossover in the normal
state of PCCO near optimal doping~\cite{fournier}.

In summary, we have studied the density of states near the fermi
level in the normal state of the electron-doped cuprate
Pr$_{2-x}$Ce$_x$CuO$_4$ ($x = 0.15$) using point contact
spectroscopy. We have observed a normal state gap in the density of
states, which persists even in a magnetic field of 31 T. The weak
magnetic field dependence of the normal state gap leads to two
possible explanations: (1) The pseudogap closing field is higher (by
about a factor of 3) than expected from a pure Zeeman relation
between $T^{*}$ and $H_{pg}$ and therefore, preformed pairing above
$T_c$ is not the origin of the pseudogap~\cite{kawakami} or (2) the
NSG observed in $n$-doped cuprates is {\em not} analogous to the
pseudogap in $p$-doped cuprates. Instead it is formed due to
electron-electron interactions at the surface of $n$-doped cuprates.

The work carried out at the NHMFL is supported by the In House
Research Program of the NHMFL which is supported by NSF cooperative
agreement No. DMR-00-84173 and by the State of Florida. The work at
Maryland was supported by the NSF under DMR-0653535.


\begin{references}
\bibitem{gvshift} The resistance of the crystal $R_{crys}$ in the point contact junction in a magnetic field $H$ is estimated
from the $G-V$ curves and the equation $R_{crys} = G_{H=0}^{-1} -
G_H^{-1}$ using the value of $G$ at 15 mV. The junction bias $V_j$
is then recalculated using the equation $V_j = V_{meas} - I \times
R_{crys}$, where $V_{meas}$ is the measured junction bias and $I$ is
the junction current. The recalculated $G-V$ curves are then shifted
to match the $H$=0 curve at $V$ = 15 mV.
\end{references}
\end{document}